\documentstyle[12pt]{article}\textheight 230mm\textwidth 160mm
\begin{document}
\pagestyle{empty}
\hspace*{11.7cm}\vspace{-3mm}KANAZAWA-94-01\\
\hspace*{13cm}March 1995

\begin{center}
{\Large\bf  Nontrivial, Asymptotically Non-free
 Gauge Theories \\and\\
Dynamical Unification of Couplings}
\end{center}

\vspace{2cm}

\begin{center}{\sc Jisuke Kubo}$\ ^{\dagger}$
\end{center}
\vspace*{0.1cm}
\begin{center}
{\em College of Liberal Arts, Kanazawa  University,
Kanazawa 920-11, Japan }
 \end{center}

\vspace*{2cm}
\begin{center}
{\sc\large Abstract}
\end{center}
\noindent
An evidence  for
 nontriviality of
asymptotically non-free (ANF) Yang-Mills theories
is found on the basis of optimized perturbation theory.
It is argued
 that  these theories with matter couplings
can be made nontrivial
by means of the reduction
of couplings, leading to
the idea of dynamical unification of couplings (DUC)
The second-order reduction of couplings in the
ANF $SU(3)$-gauged Higgs-Yukawa theory,
which is assumed to be nontrivial here,  is carried out
to motivate independent investigations on
its nontriviality and DUC.

\vspace*{4cm}
\footnoterule
\vspace*{2mm}
\noindent
$^{\dagger}$E-mail address: jik@hep.s.kanazawa-u.ac.jp

\newpage
\pagestyle{plain}
Asymptotically free (AF) theories \cite{asfree}
do not suffer from the
problem of triviality \cite{wilson}.
It is widely believed that the question
of triviality cannot be
addressed within the framework of perturbation theory, and
so far there is no real indication for
the existence of a  nontrivial
four-dimensional theory that is not AF.
It is however tempting to think that if an infrared-free theory has
an ultraviolet fixed point which is small,
 perturbation theory might be intact even near the fixed point and
hence could be applicable to the triviality problem.

About ten years ago,
Sakakibara, Stevenson, and I \cite{kubo1}
considered perturbation theory near a fixed point.
We formulated the problem as the problem of the
renormalization
scheme (RS) dependence,
because at  any finite order in  perturbation theory
even the existence of a positive zero of
the $\beta$-function depends
on RS. We performed
our investigation  on the basis of
optimized perturbation theory (OPT) \cite{stevenson1},
which as well known yields RS-invariant perturbative
 approximations and has  already experienced certain successes
in perturbative QCD \cite{qcd} and also in QED \cite{kubo4}.
We found that
one needs  a perturbative calculation
of a physical quantity
of at least third order in order to be able to apply
 our method.
Investigating concrete field theory examples,
we argued that under
certain circumstances perturbative analyses
based on  OPT near a fixed point
could be believable.

Recently,  using
the third-order QCD corrections
of the $e^{+}\,e^{-}$ cross sections \cite{surguladze1},
Mattingly and Stevenson \cite{stevenson2} applied  OPT and
concluded  that AF QCD
has an infrared-stable fixed point.
Although the assumption on the existence of an infrared
fixed point in AF QCD has no logical inconsistency,
it is not clear at all how much the fixed point
found in a perturbative approach can describe the physics
in the infrared regime, because the nonperturbative effects
play the essential role in understanding the
low energy physics of AF QCD.
In the ultraviolet regime, on the other hand,
the nonperturbative nature
may be neglected in describing the basic part of
the physics of AF QCD.

One of the main assumptions of this note,
which is partly motivated by this fact, is that this is true
even in asymptotically non-free (ANF) Yang-Mills theories.
Of course, this is a very strong assumption, but
there is neither internal inconsistency of
this assumption, nor known fact against it
(at least to my knowledge \cite{lattice}).
Moreover, as it will be seen, the investigation based on OPT
indicates that
ANF Yang-Mills theories could have an ultraviolet fixed point
so that they
could be well-defined, interacting  theories in the ultraviolet limit.

We begin by recalling the basic result obtained in Ref.\ \cite{kubo1}.
Consider
a physical quantity ${\cal R}(p_k,\mu,\alpha (\mu) /\pi)$
in a massless renormalizable theory, where
$p_k$ stand for the physical external momenta, $\mu$
is the renormalization scale, and $\alpha(\mu)$ is the
renormalized coupling.
In the {\em n}th order of perturbation theory,
 ${\cal R}$ can be written as
$ {\cal R}^{(n)}(p_k,\mu,a) ~=~\gamma\,a\,[\,1+
\sum_{i=1}^{n-1} r_{i}(p_k,\mu)
\,a^i\,]~,~ a ~\equiv~\alpha/\pi$,
while the $\beta$-function takes the form
$\beta^{(n)}(a) ~=~
-a^2\,\sum_{i=0}^{n-1}\,b_{i} a^{i}$
The coefficients $b_i$'s ($i \geq 2$) are RS-dependent, and
along with $\mu$ they  can uniquely parameterize
the RS-dependence.
Therefore,  ${\cal R}$,  being a physical quantity, has to satisfy
$\mu \partial {\cal R}/\partial \mu+
\beta \partial{\cal R}/\partial a=0$ and also
$\partial {\cal R}/\partial b_i =0~(i\geq 2)$,
which we altogether symbolically denote by
$d {\cal R}/d({\rm RS}) =0   $.
Then the essence of OPT is to demand
the optimization condition \cite{stevenson1}
\begin{eqnarray}
\left. \frac{d\, {\cal R}^{(n)}}{d\,
({\rm RS})}\, \right|_{{\rm RS}={\rm opt RS}} =0 ~,
\end{eqnarray}
and to fix from this an optimized RS for a given
physical quantity. Note that in perturbation theory one has only
$d {\cal R}^{(n)}/d({\rm RS}) =O(a^{n+1})$.
In Ref.\ \cite{kubo1}, we assumed that
 OPT makes sense even near a fixed point and found  that the
fixed point $a^{*}_{\rm opt}$ in the third order can be obtained from
\begin{eqnarray}
0 & = & \frac{7}{4}\frac{b_0}{b_1}
+a^{*}_{\rm opt}+
3 \frac{b_0}{b_1}\, \rho_2 \,(a^{*}_{\rm opt})^2~,\\
\rho_2 &=&
r_2+\frac{b_2}{b_0}-(\,r_1+\frac{1}{2} \frac{b_1}{b_0}\,)^2~,
\end{eqnarray}
where $\rho_2$
is the RS-independent quantity for a given ${\cal R}$.
{}From Eq.\ (2), one sees  that the more negative
the $\rho$'s are, the more likely is
the existence of a positive $a^{*}_{\rm opt}$.

What follows is a slight generalization of the analysis
of Ref.\ \cite{stevenson2} in QCD, but with completely different
physics and its applications in mind.
The $\beta$-function coefficients of
the first three orders in the
$\overline{{\rm MS}}$ scheme can be found in
Refs.\ \cite{tarasov1}:
\begin{eqnarray}
 b_0 &= &\frac{11}{6} C_A -\frac{2}{3} T_F f~,~
b_1=\frac{17}{12} C_{A}^{2}
 -(\,\frac{5}{6} C_A +\frac{1}{2} C_F\,)\, T_F f~,\\
b_2 &=& \frac{2857}{1728} C_{A}^{3}+(\,-\frac{1415}{864} C_{A}^{2}
+\frac{79}{432} C_A T_F f -\frac{205}{288} C_A C_F \nonumber\\
& &+\frac{11}{72} C_F T f+\frac{1}{16} C_{F}^{2}\,)\,T_F f~,
\end{eqnarray}
where $ C_A,C_F$ and $T_F$ are the usual group theoretic
coefficients.
($C_A=N,C_F = (N^2-1)/2N,T_F=1/2$
for the $SU(N)$ gauge theory with the Dirac fermions
in the fundamental representation.)
Asymptotic non-freedom requires that $f >11 C_A /2 $,
and  I concentrate
only on such cases from the reason given before.
I will below calculate $\rho_2$
in the ANF $SU(2)$ and $SU(3)$ gauge theories
with $f$ Dirac fermions in the fundamental representation.
To this end, I use the third-order
corrections to (A) $\sigma_{\rm tot} ({\rm e}^+ {\em e}^- \to
\mbox{hadrons})$ \cite{surguladze1} and (B) the Gross-Llewellyn Smith
sum rule for deep inelastic neutrino-nucleon scattering \cite{larin1}.

\vspace{0.3cm}
\noindent
A. $\sigma_{\rm tot} ({\rm e}^+ {\em e}^-
\to \mbox{hadrons})$

The first quantity is the so-called $R$-ratio
$R(s/\mu^2,a(\mu))~ =~ d_R \sum_f Q_{f}^2 \,(\,1+
\,{\cal R}(s/\mu^2,a)\,)$,
which is defined by $\sigma_{\rm tot}
({\rm e}^+ {\em e}^- \to \mbox{hadrons})/
\sigma ({\rm e}^+ {\em e}^- \to \mu^+ \mu^-)   $ in the
${\rm e}^+ {\em e}^-$ annihilation,
where $s$ is the center of mass energy, $d_R$ is the dimension
of the quark representation, and $Q_f$ stands for the electric charge
of the ``f'' quark.
Since it is unlikely that the real electric charge of
the quark is related to the existence of a fixed point
in a nonabelian gauge theory, I instead use the fermion
number and assume that $Q_f=1$ for all
the fermions.
Under this assumption, I recall the third-order result of
Ref.\ \cite{surguladze1}:
\begin{eqnarray}
{\cal R}^{(3)} &=& \frac{3}{4} C_F\, a\,(\,1+r_1 a+r_2 a^2\,)~,
\end{eqnarray}
where
\begin{eqnarray}
r_1(s/\mu^2=1) &=&
[\,\frac{41}{8}-\frac{11}{3} \zeta (3)\,] \,C_A - \frac{1}{8} C_F
 +[\,-\frac{11}{6} + \frac{4}{3} \zeta (3)\,]\, T_F f~,\\
r_2(s/\mu^2=1) &=&
[\,\frac{90445}{2592}
 - \frac{2737}{108}  \zeta (3)-
\frac{121}{432}  \pi^2 \,]\,C_{A}^{2}
 - [\,\frac{127}{48} +
\frac{143}{12}  \zeta (3)\,]\,C_A C_F -
\frac{23}{32} C_{F}^{2} \nonumber\\
& &+55 [\,-\frac{1}{18} C_A + \frac{1}{3} C_F\,]\, \zeta (5) C_A
+[\,\frac{302}{81} - \frac{76}{27} \zeta (3)-
 \frac{1}{27} \pi^2\,] \,T_{F}^{2} f^2 \nonumber\\
& &+[\,\frac{11}{144} - \frac{1}{6} \zeta (3)\,]
\frac{d^{abc}d^{abc}}{C_F d_R}f
+[\,(\,- \frac{1940}{81}+ \frac{448}{27} \zeta (3)
+\frac{10}{9} \zeta (5)+\frac{11}{54}\pi^2 \,)\, C_A \nonumber\\
& &+ (\,  - \frac{29}{48}   + \frac{19}{3} \zeta(3)  -
\frac{20}{3} \zeta (5)\, ) C_F \,]\,T_F f~.
\end{eqnarray}
Using these  three- and four-loop results,
one  can now computes $\rho_2$ defined in Eq.\ (3):
\begin{eqnarray}
\rho_2 &\simeq &
[\,-4.2140+ 0.03224 f +
    0.05455 f^2 - 8.12\times 10^{-4} f^3 \nonumber\\
& &- 1.53 \times 10^{-4} f^4\,]\cdot
[\,1- f/11\,]^{-2}~~\mbox{for} ~~SU(2) \\
 &\simeq& ~
[\,-8.4102 - 0.50203f +
    0.10845 f^2 - 2.066 \times10^{-3} f^3 \nonumber\\
& &-  6.78 \times10^{-5}f^4\,]\cdot [\,1 - 2 f/33\,]^{-2}
{}~~\mbox{for} ~~ SU(3)~,
\end{eqnarray}
where $d^{abc} d^{abc}= 0$ for $SU(2)$ and $40/3$
for $SU(3)$ have been used.
Then I investigate whether Eq.\ (2) has a positive solution
if $f >12 \,(17)$ for $SU(2)\,(SU(3))$. The result
is shown in TABLE I.

\vspace{0.1cm}
\begin{center}
TABLE I. The third-order fixed points
($\alpha_{\rm opt}^{*}=a_{\rm opt}^{*} \pi$)
 from the $R$ ratio.

\vspace{0.1cm}
\begin{tabular}{|ccc|ccc|}
\hline\hline
\multicolumn{3}{|c|}{$SU(2)$}
 & \multicolumn{3}{|c|}{$SU(3)$}\\
$f$ & $(b_0 /b_1 )\rho_2$ &
$\alpha_{\rm opt}^{*}$ &
 $f$  &  $(b_0 /b_1 )\rho_2$  & $\alpha_{\rm opt}^{*}$
\\ \hline
$12$  & $-3.317$ & $0.494$ & $17$ &
$-18.197$ & $0.096$
 \\
$13$  & $-1.912$ & $0.856$ & $18$ &
$-5.689$ & $0.294$
\\
$14$  & $-1.815$ & $0.960$ & $19$ &
$-3.794$ & $0.441$
\\
$15$  & $-2.014$ & $0.940$ & $20$ &
$-3.365$ & $0.516$
\\ \hline\hline
\end{tabular}

\end{center}

\vspace{0.1cm}
\noindent
As one can see from TABLE I, $\alpha_{\rm opt}^{*}$
for some cases is small so that one may
trust the  results.

\vspace{0.3cm}
\noindent
B. The Gross-Llewellyn Smith sum rule

This sum rule says that
the first moment of the isospin singlet structure function
 for the hadronic matrix element
which describes deep inelastic processes is six
at the parton model level;
$\int_{0}^{1} dx (\,F_{3}^{\overline{\nu} p}
+F_{3}^{\nu p}\,)(x,Q^2/\mu^2,a)
{}~=~ 6\,(\,1+{\cal R}(Q^2/\mu^2,a)\,)$,
where $x$ is one of the scaling variables
 in the processes.
The third-order QCD correction
has been computed by Larin and Vermaseren \cite{larin1}:
 \begin{eqnarray}
{\cal R}^{(3)} &=& \frac{3}{4} C_F \,a\,(\,
1+r_1 a+r_2 a^2\,)~,\\
r_1(Q^2/\mu^2=1) &=&
\frac{23}{12} C_A - \frac{7}{8} C_F- \frac{1}{3} f~,\\
r_2(Q^2/\mu^2=1) &=&
[\,\frac{5437}{648}  - \frac{55}{18} \zeta (5)]\, C_{A}^{2}
- [\frac{1241}{432} - \frac{11}{9} \zeta (3)\,] \,C_A C_F
+ \frac{1}{32} C_{F}^{2} \nonumber\\
 & &+[\,(\,-\frac{3535}{1296} - \frac{1}{2} \zeta (3)+
   \frac{5}{9} \zeta (5)\,)\, C_A
+ (\,\frac{133}{864} +\frac{5}{18} \zeta (3)\,)\,C_F \nonumber\\
& &+ (\,\frac{11}{144}-\frac{1}{6} \zeta (3)\,)\,
\frac{d^{abc}d^{abc}}{C_F N_C} \,] \,f
+\frac{115}{648} f^2~.
\end{eqnarray}
As in the case A, I insert the $r_1$ and $r_2$ into the r.h.side
of Eq.\ (3) and
obtain
\begin{eqnarray}
\rho_2 &\simeq&
[\,6.8068  - 3.90512 f +
   0.57496f^2-3.157\times 10^{-2} f^3 \nonumber\\
& &+ 5.48 \times10^{-4} f^4\,]\cdot
[\,1-  f/11\,]^{-2}~~\mbox{for}~~SU(2)~ \\
&\simeq&
[\,16.5809 - 6.45245 f +
    0.630222 f^2 - 2.2537\times 10^{-2} f^3 \nonumber\\
& &+ 2.44\times 10^{-4} f^4\,]\cdot
[\,1 - 2 f/33\,]^{-2}~~\mbox{for}~~SU(3)~.
\end{eqnarray}
The values of $\rho_2$ and $\alpha_{\rm opt}^{*}$
for some different $f (> 11 C_A /2)$ are shown in TABLE II.

\vspace{0.1cm}
\begin{center}
TABLE II. The third-order fixed points
($\alpha_{\rm opt}^{*}=a_{\rm opt}^{*} \pi$)
from the Gross-Llewellyn Smith sum rule.

\vspace{0.1cm}
\begin{tabular}{|ccc|ccc|}
\hline\hline
\multicolumn{3}{|c|}{$SU(2)$}
 & \multicolumn{3}{|c|}{$SU(3)$}\\
$f$ & $(b_0 /b_1)\rho_2$ &
$\alpha_{\rm opt}^{*}$ &
 $f$  &  $(b_0 /b_1)\rho_2$  & $\alpha_{\rm opt}^{*}$
\\ \hline
$12$  & $-2.896$ & $0.568$ & $17$ &
$-17.196 $ & $0.100$
 \\
$13$  & $-1.279$ & $1.133$ & $18$ &
$-4.681$ & $0.339$
\\
$14$  & $-1.063$ & $1.333$ & $19$ &
$-2.766$ & $0.558$
\\
$15$  & $-1.104$ & $1.314$ & $20$ &
$-2.296$ & $0.684$
\\ \hline\hline
\end{tabular}
\end{center}

\vspace{0.1cm}
\noindent
The results are surprisingly similar to
those for A. This again supports the reliability of
the fixed point analysis based on OPT, and may be seen as
 an evidence for
 ultraviolet fixed points
in the ANF Yang-Mills theories.

Triviality of gauged Higgs-Yukawa systems is
widely expected, unless they are completely asymptotically free.
A rigorous treatment of
the  asymptotic behavior
of theory with more than one couplings is given in Ref.\
\cite{zim1}. It was found \cite{kubo2} that
 by imposing  a certain
relation among the gauge, Higgs, and Yukawa couplings
which are consistent with perturbative renormalizability,
it is possible to make
the $SU(3)$-gauged Higgs-Yukawa system
completely  asymptotically free
and hence nontrivial \cite{kugo}.
This renormalization group invariant
relation among couplings
is a consequence of the ``reduction of couplings'' \cite{zim1}.

Inspired by the possibility that
ANF Yang-Mills gauge theories may
be nontrivial under certain circumstances
and by the fact that gauged Higgs-Yukawa systems can be made
asymptotically free by means of the
reduction of couplings,
one may be naturally led to the idea that
even ANF gauged Higgs-Yukawa
systems are nontrivial if the reduction of
couplings is appropriately carried out.
One then would achieve a {\em dynamical
gauge-Higgs-Yukawa unification} in a theory, because these
couplings are forced
in a dynamically consistent fashion to be related
with each other in order
for the theory to remain  well-defined and
interacting in the ultraviolet limit.

OPT for systems with more than one couplings does not exist yet,
because there is no known systematic way how to control the
propagation of the RS-dependence of lower orders
to higher orders.
But it is  clear that once the reduction of couplings is
applied to a system with many couplings so that the reduced system
contains only one independent coupling, one
can employ all the facilities of  OPT.
Unfortunately,
third-order  calculations
in gauged
Higgs-Yukawa systems do not exist yet.
Here I would like to
present the result of the two-loop reduction in
the ANF $SU(3)$-gauged Higgs-Yukawa
theory  to motivate corresponding
 higher order calculations.

Let me first mention few words about the reduction of couplings,
and consider a massless, renormalizable
gauge theory based on a simple
gauge group with $N$ other couplings, where
the gauge coupling is denoted by $\alpha$,
and the others by $\alpha_i~,~i=1,\cdots,N$.
The complete reduction of couplings \cite{zim1} is
equivalent to demand that $\alpha_i $ be written as a power series
of $\alpha$, i.e.,
$\alpha_{i} ~=~
\sum_{n=0}^{\infty}\eta^{(n)}_{i}\,(\alpha/\pi)^{n}\,a~,~
i=1,\cdots,N$.
As the consequence, the reduced system contains only $\alpha$ as
the independent coupling--unification of couplings.
 It was shown \cite{zim1} that
the power series
is consistent with perturbative renormalizability
only if the reduction equations
\begin{eqnarray}
\beta_{\alpha}(\alpha,\alpha_i(\alpha))
 \,\frac{d \alpha_{i}(\alpha)}{d \alpha}
&=&\beta_{i}(\alpha,\alpha_i(\alpha))
\end{eqnarray}
are satisfied, where $\beta_{\alpha}$ stands for
 the $\beta$-function of $\alpha$, and
$\beta_{i}$ for
that of $\alpha_{i}$.
The uniqueness of the power series solution
can be decided at the one-loop level, and
the $\eta$'s can be computed order by order in perturbation theory
\cite{zim1}.

The gauged Higgs-Yukawa model I consider below
can be obtained from the standard model by
switching off the $SU(2)$ and $U(1)$ gauge couplings,
dropping all leptons, and allowing $n_d$ families of quarks.
I also assume that
only one of the
(up-type) Yukawa couplings is nonvanishing;
the simplified system contains only
the $SU(3)$ gauge coupling $\alpha$, the Yukawa coupling
$\alpha_t$, and the Higgs self-coupling $\alpha_{h}$.
(For $n_d \leq 8$, this system can be made completely
asymptotically free \cite{kubo2}.) Here I am interested in the case
for  $n_d > 8$, and recall the $\beta$-functions \cite{beta2}
\begin{eqnarray}
\frac{\beta_3}{\pi} &=&
a^2\,[\,-\frac{11}{2} + \frac{2}{3} n_d+
(\, \frac{19}{6} n_d - \frac{51}{4}\,) a
- \frac{1}{4} a_t+\cdots\,]~,\\
\frac{\beta_t}{\pi} &=&
a_t\,[\,\frac{9}{4} a_t-4 a+
  \frac{9}{2} a a_t - \frac{3}{4} a_h a_t - \frac{3}{2} a_{t}^{2}
+(\,\frac{10}{9} n_d - \frac{101}{6}\,) a^2 +
\frac{3}{16} a_{h}^{2}+\cdots\,]~,\\
\frac{\beta_h}{\pi} &=&
3 a_{h}^{2} + 3 a_h a_t - 3 a_{t}^{2}
 - 4 a a_{t}^{2} \nonumber\\
& &- \frac{3}{16} a_h a_{t}^{2} + \frac{15}{4}a_{t}^{3}
- \frac{39}{8} a_{h}^{3} + 5 a a_h a_t
- \frac{9}{2} a_{h}^{2} a_t+\cdots~,
\end{eqnarray}
where $a_i=\alpha_{i}/\pi$.
It can be shown that
the power series solution of the reduction equations (16)
with $i=t~,~h$, i.e.,
$$\alpha_{i} ~=~
\sum_{n=0}^{\infty}\eta^{(n)}_{i}
\,(\frac{\alpha}{\pi})^{n}\,a~,~
i=t,h~,$$
exists uniquely to all orders in perturbation theory
so that the original system
with three independent couplings can uniquely be reduced
to a system with only one independent coupling, $\alpha$.
 The first- and second-order coefficients can be computed
by solving Eq.\ (16) with
 the second-order $\beta$-function (17)-(19), and the results are
 given in TABLE III.

\begin{center}
TABLE III. The expansion coefficients for the reduction of couplings
in the $SU(3)$-gauged Higgs-Yukawa theory.

\vspace{0.1cm}
\begin{tabular}{|ccccc|}
\hline\hline
$n_d$
 & $\eta^{(0)}_{t}$ &
$\eta^{(1)}_{t}/\pi$ & $\eta^{(0)}_{h}$ &
$\eta^{(1)}_{h}/\pi$\\
$9$  & $2$ & $3.294$ & $1.283$ &
$2.586$
 \\
$10$  & $2.296$ & $4.356$ & $1.533$ &
$3.592$
\\ \hline\hline
\end{tabular}
\end{center}

\vspace{0.1cm}
\noindent
The reduced system has only one $\beta$-function
$$\frac{\beta}{\pi} ~=~a^2\,[\,
-\frac{11}{2}+\frac{2}{3} n_d+
(\,-\frac{151}{12}+\frac{169}{54} n_d\,)a+O(a^2)\,]~.$$
The fact that the first two coefficients of
$\beta$ for $n_d \geq 9$ are
positive (as they are in the previous cases)
does not mean anything about a fixed point
within the framework of OPT;
one needs a complete third-order calculation
to obtain $\rho_2$ and then to solve Eq.\ (2).
If it will be negative and large, there will be
a small, positive $a^{*}_{\rm opt}$.

There will be many applications of
the idea of dynamical unification of couplings (DUC)
in constructing realistic unified gauge models.
Unification of the gauge couplings in
 ANF extensions of the standard
model, for instance,  were previously considered in Refs.\
\cite{maiani}.
In contrast to the present idea, it was assumed there
that
 the gauge couplings asymptotically diverge
 so that  if one requires
the couplings to become strong
simultaneously at a certain energy scale,
one can predict their low energy values \cite{bardeen}.
There are many papers based on this idea,
but non of them discusses
  nontriviality of ANF unified gauge models
and its possible relation to unification of couplings.
Obviously, it is desirable to justify
the assumptions (specified in the text) leading to the idea
 of DUC
independently in different approaches.

I would like to thank T. Kugo
for stimulating discussions,
which  led me to
consider DUC
in ANF theories, and
 also G. Schierholz and T. Suzuki
for useful information. I am greatly indebted to
W. Zimmermann for continuos supports and encouragement.


\begin{thebibliography}{99}

\bibitem{asfree}G.\ 't Hooft, unpublished (1972);
 D.\ J.\ Gross and F.\ Wilczek, Phys.\ Rev.\ Lett.\ {\bf 30}, 1343
(1973);
 H.\ D.\ Politzer, Phys.\ Rev.\ Lett.\ {\bf 30}, 1346 (1973).


\bibitem{wilson}K.\ Wilson, Phys.\ Rev.\ D {\bf 6}, 419 (1972);
K.\ Wilson and J.\ Kogut,  Phys.\ Rep.\ {\bf 12C}, 75 (1974);
B.\ Freedman, P.\ Smolensky, and D.\ Weingarten,
Phys.\ Lett.\ B {\bf 113}, 481 (1982);
J.\ Fr{\" o}hlich, Nucl.\ Phys.\
{\bf B200} [FS4], 281 (1982);
D.\ J.\ E.\ Callaway, Phys.\
Rep.\ {\bf 167}, 241 (1988) and references therein.

\bibitem{kubo1} J.\ Kubo, S.\ Sakakibara, and P.\ M.\ Stevenson,
Phys.\ Rev.\ D {\bf 29}, 1682  (1984).


\bibitem{stevenson1} P.\ M.\ Stevenson,
Phys.\ Rev.\ D {\bf 23}, 2916  (1981); Nucl.\ Phys.\ {\bf B 231},
65 (1984).

\bibitem{qcd}See, e.g., P.\ Aurence, R.\ Baier, M.\ Fontannaz,
Z.\ Phys.\ C {\bf 48}, 143 (1990).


\bibitem{kubo4} J.\ Kubo and S.\ Sakakibara,
Z.\ Phys.\ C {\bf 14}, 345 (1982).

\bibitem{surguladze1}
L.\ R.\ Surguladze and M.\ A.\ Samuel, Phys.\ Rev.\ Lett.\
{\bf 66}, 560 (1991);
S.\ G.\ Gorishny, A.\ L.\ Kataev, and S.\ A.\ Larin, Phys.\ Lett.\
B {\bf 259}, 144 (1991)
and references therein for the calculations
 in lower orders.


\bibitem{stevenson2} A.\ C.\ Mattingly and P.\ M.\ Stevenson,
Phys.\ Rev.\ Lett.\ {\bf 69}, 1320 (1992);
Phys.\ Rev.\ D {\bf 49}, 437  (1994).

\bibitem{lattice}
QCD with many flavors has been recently investigated in:
Y.\ Iwasaki, K.\ Kanaya, S.\ Sakai, and T.\ Yoshie,
Phys.\ Rev.\ Lett.\ {\bf 69}, 21 (1992);
Nucl.\ Phys.\ B (Proc. Suppl.) {\bf 34}, 314 (1994).
They found that
nonperturbative phenomena such as quark confinement
disappear with increasing number of flavors.


\bibitem{tarasov1}
O.\ V.\ Tarasov, A.\ A.\ Vladimirnov, and A.\ Yu.\ Zharkov,
Phys.\ Lett.\ B {\bf 83}, 429 (1980);
S.\ A.\ Larin and J.\ A.\ M.\ Vermaseren,
Phys.\ Lett.\ B {\bf 303}, 334 (1993)
and references therein for the calculations
of the  $\beta$-functions in lower orders.

\bibitem{larin1}
S.\ A.\ Larin and J.\ A.\ M.\ Vermaseren,
Phys.\ Lett.\ B {\bf 259}, 345 (1991)
and references therein for the calculations
 in lower orders.

\bibitem{zim1}W.\ Zimmermann, Commun.\ Math.\ Phys.\ {\bf 97}, 211
(1985);
R.\ Oehme and W.\ Zimmermann, Commun.\ Math.\ Phys.\ {\bf 97}, 569
(1985);  R.\ Oehme, Prog.\ Theor.\ Phys.\ Suppl.\
{\bf 86}, 215 (1986).

\bibitem{kubo2} J.\ Kubo, K.\ Sibold, and W.\ Zimmermann,
Nucl.\ Phys.\ {\bf B259}, 331 (1985).

\bibitem{kugo}Asymptotic freedom of gauged Higgs-Yukawa systems
may be closely related to the nonperturbative existence
of gauged
Nambu-Jona-Lasino models. See
T.\ Kugo, Prog.\ Theor.\ Phys.\ {\bf 92}, 1161 (1994),
and also
K.\ -I.\ Kondo, M.\ Tanabashi, and K.\ Yamawaki,
Prog.\ Theor.\ Phys.\ {\bf 89}, 1249 (1993);
K.\ -I.\ Kondo, A.\ Shibata, M.\ Tanabashi, and K.\ Yamawaki,
Prog.\ Theor.\ Phys.\ {\bf 91}, 541 (1994).

\bibitem{maiani}
L.\ Maiani, G.\ Parisi, and R.\ Petronzio,
Nucl.\ Phys.\ {\bf B136},
115 (1978); N.\ Cabibo and G.\ R.\ Farrar, Phys.\ Lett.\ B {\bf 110},
107 (1982);
 L.\ Maiani and R.\ Petronzio,
Phys.\ Lett.\ B {\bf 176}, 120 (1986).
See also
G. Grunberg, Phys.\ Rev.\ Lett.\ {\bf 58}, 1180 (1987)
and
J.\ Kubo, M.\ Mondragon, N.\ D.\ Tracas, and G.\ Zoupanos,
Phys.\ Lett.\ B {\bf 342}, 155 (1995)
for approaches related to the present one.

\bibitem{bardeen}
Diverging of certain couplings in the standard model
has been interpreted as the compositeness condition
of the Higgs particle in:
W.\ Bardeen, C.\ Hill, and M.\ Lindner,
Phys.\ Rev.\ D {\bf 41}, 1647 (1990).

\bibitem{beta2}
 V.\ Barger, M.\ S.\ Berger, and P.\ Ohmann,
Phys.\ Rev.\ D {\bf 47}, 1093 (1993) and references therein.

\end{thebibliography}
\end{document}